\documentstyle[12pt]{article}
\begin{document}
\topmargin=9.5mm\oddsidemargin=9.5mm\textwidth=140mm 
\def\baselinestretch{1.5}
\newcommand{\beq}{\begin{equation}}
\newcommand{\eeq}{\end{equation}}
\newcommand{\beqn}{\begin{eqnarray}}
\newcommand{\eeqn}{\end{eqnarray}}
\newcommand{\dpf}{\displaystyle\frac}
\newcommand{\no}{\nonumber}
\begin{center}
{\Large Geometry of the extreme Kerr black hole }
\end{center}
\vspace{1ex}
\centerline{\large Bin Wang$^{1,2}$\footnote[0]{email:binwang@fudan.ac.cn},
    \ Ru-Keng Su$^{2,1}$\footnote[0]{email:rksu@fudan.ac.cn},\
P.K.N.Yu$^{3}$}
\begin{center}
{$^1$ Department of Physics, Fudan University, Shanghai 200433, P.R.China\\
$^2$ China Center of Advanced Science and Technology (World Laboratory),
 P.O.Box 8730, Beijing 100080, P.R.China\\
$^3$ Department of Physics and Materials Science, City University of
Hongkong,
 Hongkong, P.R.China}
\end{center}
\vspace{6ex}
\begin{abstract}
Geometrical properties of the extreme Kerr black holes in the topological
sectors of nonextreme and extreme configurations are studied. We find that
the Euler characteristic plays an essential role to distinguish these two
kinds of extreme black holes. The relationship between the geometrical
properties and the intrinsic thermodynamics are investigated.
 \end{abstract}
 \vspace{6ex}
 \hspace*{0mm} PACS number(s): 04.70.Dy, 04.20.Gz, 04.62.+v
 \vfill
 \newpage

I. Introduction

The entropy of the extreme black hole (EBH) has been an intriguing subject
of investigations,
recently. Based upon the basic difference between the topology of the EBH
and
non-extreme black hole (NEBH), Hawking et al.[1,2] claimed that the EBH is
a different
object from its non-extreme counterpart and the Bekenstein-Hawking formula
of the entropy fails to
describe the entropy of EBH. The EBH must have zero entropy, despite the
non-zero
area of the event horizon, and can be in thermal equilibrium at arbitrary
temperature.

On the other hand, starting from grand canonical ensemble, Zaslavskii [3]
argued
that a black hole can approach the extreme state as closely as one likes in
the topological
sector of non-extreme configurations. In so doing, the thermodynamic
equilibrium can be fulfilled
at every stage of the limiting process and the Bekenstein-Hawking formula
of entropy is still valid for the final EBH.
To study the geometry of non-extreme Reissner-Nordstrom (RN) black hole
near the extreme state, Zaslavskii [4] found
that the limiting geometry of the RN black hole depends only on one scale
factor and
the whole Euclidean manifold is described by the Bertotti-Robinson(BR)
spacetime.

The above contradiction seems to imply that the geometric properties, in
particular,
the spacetime topology, play an essential role in the explanation of
intrinsic
thermodynamics of the extreme black holes. To exhibit the relationship
between
the topology and the thermodynamical features of gravitational instantons,
many authors [5,6] introduced
the Euler characteristic in the new formulation of entropy and found that
the Euler
characteristic determines the entropy of NEBH directly. They also found
that if
the EBH satisfies the topological requirment of Hawking et al[1], the Euler
characteristic
is zero [5]. But for the limiting metric of NEBH near the extreme state
suggested
by Zaslavskii, the Euler characteristic has not been calculated and the
relationship
between the entropy and Euler characteristic for BR metric has not been
addressed.

In this paper we hope to extend the results of RN black hole to rotating
Kerr black hole. We
will focus our attentions on the extreme Kerr black hole obtained in the
grand canonical ensemble and
compare its Euler characteristic, thermodynamical behaviors with that of
the original Kerr EBH.
We will show the geometrical properties as well as the thermodynamical
properties
of the Kerr EBH in the topological sectors of non-extreme and extreme
configurations
are quite different. Since the extreme conditions are both satisfied for
these
two configurations, we speculate that perhaps there are two kinds of EBH in
the nature.

The organization of this paper is as follows: in Sec.II and Sec.III, we
present
the geometric properties of the Kerr EBH in the topological sectors of
non-extreme and extreme
configurations respectively. Sec.IV is devoted to the calculation of the
thermodynamical
quantities. The discussions and conclusions will be presented in the last
section.
\newpage
II. Extreme black hole with non-extreme topology

The metric of the Kerr black hole reads[7]:
 \beqn                               
 {\rm d}s^2&=&-\dpf{\triangle}{\Sigma}[{\rm d}t-a\sin^2\theta{\rm d}\phi]^2
 +\dpf{\sin^2\theta}{\Sigma}[(r^2+a^2){\rm d}\phi-a{\rm d}t]^2 \no \\
           & &+\dpf{\Sigma}{\triangle}{\rm d}r^2 +\Sigma{\rm d}\theta^2
\eeqn
where
\beq             
        \triangle =r^2-2Mr+a^2,    \\  a=J/M
\eeq
\beq    
        \Sigma=r^2+a^2\cos^2\theta
\eeq
$J$ and $M$ are respectively the angular momentum and the mass of the Kerr
black hole.
It displays an event horizon and Cauchy horizon provided that $a^2< M^2$
and locate at $r_+=M+\sqrt{M^2-a^2}$ and $r_-=M-\sqrt{M^2-a^2}$
respectively.
For the extreme case $M=a$, these two horizons degenerate and only event
horizon $r_+=M$
exists.

As we will be interested in the metric near the horizon $r_+$, it is
convenient
to redefine the angular variable [4] according to
\beq              
{\rm d}\phi={\rm d}\phi+\dpf{a-a\sqrt{f}}{r^2+a^2}{\rm d}t
\eeq
where
\beq        
f=\dpf{\triangle}{r^2+a^2}=\dpf{(r-r_+)(r-r_-)}{r^2+a^2}
\eeq
The metric (1) can be rewritten as
\beqn              
{\rm d}s^2 & = & -\dpf{r^2+a^2}{\Sigma}[\sqrt{f}{\rm
d}t-a\sin^2\theta\sqrt{f}{\rm d}\phi
   -\sqrt{f}\dpf{a^2\sin^2\theta(1-\sqrt{f})}{r^2+a^2}{\rm d}t]^2 \no \\
           &   & +\dpf{\sin^2\theta}{\Sigma}[(r^2+a^2){\rm
d}\phi-a\sqrt{f}{\rm d}t]^2+\dpf{\Sigma}{r^2+a^2}{\rm d}l^2+\Sigma{\rm
d}\theta^2
\eeqn
where $l$ is the proper distance between $r_+$ and $r$.

Following the general approach for finite-size thermodynamics [8], we
consider the
grand-canonical ensemble and put the hole into a cavity. The boundary of
the cavity is
$r_B$. For the spacetime[3], the equilibrium condition reads
\beq            
\beta=\beta_0[f(r_B)]^{1/2}, \no \\
T_0=T_H=\dpf{f'(r_+)}{4\pi}
\eeq
As in ref[4], we normalize the time by the condition $t_1=2\pi T_0t$ and
choose the coordinate
according to
\beq              
r-r_+=4\pi T_0b^{-1}\sinh^2\dpf{x}{2},  b=f''(r_+)/2
\eeq
In the limit $r_+\rightarrow r_B$, where the hole tends to occupy the
entire cavity, the
region $r_+\leq r\leq r_B$ shrinks and we can expend $f(r)$ in a power
series $f(r)=4\pi T_0(r-r_+)+b(r-r_+)^2+\cdots$ near $r=r_+$. After
substituting
Eqs(7,8) into (6) and taking the extremal limit $r_+=r_-=r_B$, we obtain
\beq
{\rm d}s^2=\Sigma_B[-\sinh^2 x{\rm d}t_1^2+{\rm d}x^2+{\rm d}\theta^2]
           +\dpf{\sin^2\theta}{\Sigma_B}[(r_B^2+a^2){\rm d}\phi-a\sinh
x\sqrt{r_B^2+a^2}{\rm d}t_1]^2
\eeq
where
\beq
\Sigma_B=r_B^2+a^2\cos^2\theta,  {\rm d}x^2=\dpf{{\rm d}l^2}{r_B^2+a^2}
\eeq

This is an extension of the Bertotti-Robinson (BR) spacetime [9] metric to
the case
of the limiting form of rotating four-dimensional Kerr black hole. It can
easily be
seen that this metric has the properties of BR spacetime, namely,
nonsingular and
static [10]. This is the asymptotic form of the metric and fields near the
extremal Kerr
black hole horizon. Extending this spacetime to Kerr-Newman black hole is
straightforward.

Now we are in a position to discuss the properties of Eq(9). The horizon of
the black hole is
determined by
\beq        
\triangle=f(r_B^2+a^2)=0
\eeq
In the extreme case $T_0=T_H=\dpf{f'(r_+)}{4\pi}=0$, therefore Eq(11) can
be written as
\beq        
\triangle=(r_B^2+a^2)\dpf{f'^{2} (r_+)}{4}(b^{-1}\sinh^2
x)=(r_B^2+a^2)^2\dpf{f'^{2} (r_+)}{4}\sinh^2 x=0
\eeq
So the horizon can locate at finite $x$, say $x=0$. The proper distance
between the horizon and
any other point is finite.

It is of interest to study the topology of this extreme Kerr black hole.
Since this EBH is obtained by first taking the boundary condition
$r_+\rightarrow r_B$, and
then adopting the extremal limit, the formula of the Euler
characteristic[6] can be used directly.
We obtain
\beqn                    
\chi & = & \dpf{Mr_+(r_+ -M)}{4\pi^2}\int^{\beta_0}_0 dt\int^{2\pi}_0
d\phi\int^{\pi}_0\dpf{(r_+^2-3a^4\cos^4\theta)}{(r_+^2+a^2\cos^2\theta)^3}
\sin^2\theta d\theta\mid_{extr} \no \\
     & = & \dpf{2}{\pi} {\beta}_0 (r_+ -M)
\dpf{Mr_+}{(r_+^2+a^2)^2}\mid_{extr}
\eeqn
Taking account of
\beq              
\beta_0=\dpf{1}{T_0}=\dpf{4\pi Mr_+}{\sqrt{M^2-a^2}}
\eeq
We find that even in the extreme case $(M=a)$, $\chi=2$. This result is the
same as that of the non-extreme Kerr black hole. Therefore if we
first take the boundary condition and then let the hole become extreme, we
obtain the final
extreme Kerr black hole is still in the topological sector of nonextreme
configuration.

III. Extreme black hole with extreme topology

Now we turn to concentrate our attention on the original extreme Kerr black
hole.
This black hole satisfy $M=a$ from the very beginning. We put it in a
cavity with boundary
$r_B$. The metric has the form
\beq            
{\rm d}s^2=-\dpf{(r-r_+)^2}{\Sigma}[{\rm d}t-a\sin^2\theta{\rm d}\phi]^2
+\dpf{\sin^2\theta}{\Sigma}[(r^2+a^2){\rm d}\phi-a{\rm
d}t]^2+\dpf{\Sigma}{(r-r_+)^2}{\rm d}r^2+\Sigma{\rm d}\theta^2
\eeq
where $f=\dpf{(r-r_+)^2}{r^2+a^2}$ now. Expanding the metric coefficients
near
$r=r_+$ and introducing $r-r_+=r_B\rho^{-1}$[4], one obtains,
\beqn                   
{\rm d}s^2 & = & \Sigma_B\rho^{-2}\{-\dpf{r_B^2}{\Sigma^2_B}[{\rm
d}t-a\sin^2\theta{\rm
d}\phi]^2+\dpf{\rho^2\sin^2\theta}{\Sigma^2_B}[(r_B^2+a^2){\rm d}\phi
-a{\rm d}t]^2 \no \\
           &   & +{\rm d}\rho^2+\rho^2{\rm d}\theta^2\}
\eeqn
in the limit $r_+\rightarrow r_B$.

By using
\beq              
\triangle=(r_B^2+a^2)f=(r_B^2+a^2)\dpf{r_B^2\rho^{-2}}{r_B^2+a^2}=r_B^2\rho^
{-2}=0
\eeq
to determine the horizon, we find that the horizon locates at infinity
$\rho=\infty$.
So the distance between the horizon and any other $\rho<\infty$ is
infinite. It is this property
that gives rise to the qualitatively different topological feature of this
black hole from that
of Sec.II and plays an important role to determine its Euler characteristic
and entropy.
The metric of the hole with infinite proper distance does not show any
conical
structure near its event horizon, so no conical singularity removal is
required. It means that
$\beta_0$ can not be fixed. Applying the argument in [5,2], this feature
will
lead unambiguously to $\chi=0$ for the original extreme Kerr black hole.
The topology
of the original extreme Kerr black hole differs greatly from the extreme
Kerr black hole
obtained from its nonextremal counterpart in the grand canonical ensemble.

IV. Thermodynamical properties

By means of the relation between the Euler characteristic and the entropy
derived in [6]
\beq      
S=\dpf{A}{8}\chi
\eeq
and the different Euler characteristic obtained in Secs.II and III,
naturally one can conclude that
the extreme Kerr black hole with non-extreme topology has the entropy of
$A/4$, while
for the black hole with extreme topology zero entropy emerges. These
results can also be
deduced from the direct thermodynamic study discussed below.

We focus our attention on the extreme Kerr black hole developed from its
nonextreme counterpart
discussed in Sec.II first.

The temperature on the boundary of the cavity is $T=1/\beta$. In the grand
canonical ensemble, only this temperature
has physical meaning. The condition of thermal equilibrium has the form of
Eq(7).
By setting $r_+\rightarrow r_B$ at first and imposing the extreme condition
afterwards
\beq         
\beta=2\pi\sinh x_B\sqrt{r_B^2+a^2}
\eeq
The finite $\beta$ here is similar to that in the RN case[3]. Therefore
there exists
a well defined, in thermodynamical sense, extreme Kerr black hole state of
its
non-extreme counterpart in a grand canonical ensemble.

The action for Kerr black hole derived in [8] has the form
\beq                    
I=-\dpf{1}{4}A_H+\oint_B
d^2x\sqrt{\sigma}[\beta\dpf{dE}{dA}-(\beta\omega)\dpf{dJ}{dA}]
\eeq
where $E=\dpf{1}{8\pi}\oint_Bd^2x\sqrt{\sigma}(k-k^0)$. $k$ is the
extrinsic curvature of
the boundary embedded into two-dimensional space. $k^0$ is a constant and
can
be chosen to zero to normalize $E=0$ in a flat spacetime. Choosing the
boundary
$B$ as an isothermal surface, then the energy term in $I$ becomes $(\beta
E)_B=constant$ [8].

The free energy
\beq     
F=\dpf{I}{\beta}=-\dpf{1}{4}A_H
T+\oint_Bd^2x\sqrt{\sigma}[\dpf{dE_B}{dA}-\omega \dpf{dJ}{dA}]
\eeq
For the extreme Kerr black hole developed from the non-extremal Kerr black
hole in the grand
canonical ensemble, $T\ne 0$. Using the formula $S=-(\dpf{\partial
F}{\partial T})_D$,
where $D$ indicates the thermal quantity, we have
\beq          
S=\dpf{A}{4}
\eeq

But for the original extreme Kerr black hole, even if one let
$r_+\rightarrow r_B$
in the end,
\beq          
\beta=\dpf{4\pi r_B}{f'(r_+)\sqrt{r_B^2+a^2}\rho_B}
\eeq
On the cavity $\rho_B$ is finite, therefore $\beta$ still diverges because
of $1/f'(r_+)$.
The temperature detected on the cavity boundary for the original extreme
Kerr black hole is zero. Directly using the approach of [8] and Eq(21), we
have
\beq           
F=\oint_B d^2x \sqrt{\sigma}[\dpf{dE_B}{dA}-\omega\dpf{dJ}{dA}]
\eeq
We note that the free energy is dependent on only the thermodynamic
quantity, namely $J$,
therefore
\beq           
S=0
\eeq

These results are in consistent with the different topological properties
of these
extreme Kerr black holes.

V. Conclusions and discussions

In this paper we have studied the geometrical properties of the extreme
Kerr black hole
developed from the non-extreme one and the extreme Kerr black hole at the
very beginning
respectively. We have shown that there exists the extreme state of
non-extreme
Kerr black hole which has the universal form of the limiting metric. From
this limiting
metric and the Euler characteristic, we found that these two kinds of
extreme Kerr
black holes are in the different topological sectors, say nonextreme and
extreme
configurations, respectively. And due to the differences in the spacetime
topology
of these two kinds of extreme Kerr black holes, the intrinsic
thermodynamical
properties are quite different. The result obtained here is an extension of
that of the
spherically symmetrical RN black holes.

Combining the spherical results given in [1-4] and the nonspherical
rotating
results got above, we have an impression that there are two kinds of
extreme black holes
which have different topologies ($\chi=2$ or $zero$) and different
thermodynamical properties,
($S=A/4$ or $zero$) in the nature.

This conclusion affects not only the understanding of the geometry and
thermodynamics
of EBH, but also the phase transition of the black hole. It has been shown
by many
authors [11-14] that a phase transition exists for the Kerr black holes at
the
extreme limit. The transitional point is a critical point and the critical
behavior can
be described by various critical components satisfying the scaling law[13].
As was argued by Hawking et al [1], one should regard NEBH and EBH with
extreme topological
configuration ($\chi=0$) as qualitatively different objects and a NEBH
cannot be turned
to this kind of EBH. But as shown in Secs.II and IV, a NEBH can be
transformed
to an EBH with nonextreme topological configuration ($\chi=2$). At the
extreme
limit, a phase transition happens. Since the entropy changes continuously
from NEBH to EBH with $\chi=2$,
we come to a conclusion that this is a second order phase transition. This
result is in
consistent with previous studies [11,13].

It is widely believed that black holes retain only very limited information
about
the matter that collapsed to form them. This information is reflected in
the number of parameters
characterizing the black hole. For Kerr black hole, such parameters are the
mass
$M$ and angular momentum $J$, and the properties of the hole are completely
determined
by these parameters. This is known as the ``no hair theorem". While the
spherical results
in[1-4] and the nonspherical results obtained in our paper suggested that
the no hair
theorem is violated. A topological hair should be introduced, at least in
the extreme cases,
to describe two kinds of extreme black hole with profoundly different
properties.
This is another challenge to the ``no hair theorem" besides those proposed
in [15,16].

\end{document}